\documentclass[12pt]{article}%

\usepackage{amsmath,amssymb,amsthm,enumerate,color,graphics,epsfig,url}

\usepackage[pdftex,
colorlinks,%
linkcolor=blue,citecolor=red,urlcolor=blue,
hyperindex,%
plainpages=false,%
bookmarksopen,%
bookmarksnumbered,%
unicode]{hyperref}

\newcommand{\doi}[1]{\href{https://doi.org/#1}{doi:\nolinkurl{#1}}}
\newcommand{\arxiv}[1]{\href{https://arxiv.org/abs/#1}{arXiv:#1}}

\newtheorem{theorem}{Theorem}

\theoremstyle{definition}
\newtheorem*{problem}{Problem}
\newtheorem{remark}{Remark}

\begin{document}
\par\noindent{\LARGE\bfseries
Group classification and equivalence\\ of two classes of evolution equations\\ of arbitrary order
\par}

\vspace{5mm}\par\noindent{\large
Vyacheslav M.\ Boyko$^{\dag\ddag}$ and Oleksandra V.\ Lokaziuk$^{\dag\S}$
}

\vspace{4mm}\par\noindent{\small\itshape
$^\dag$~Institute of Mathematics of NAS of Ukraine, Kyiv, Ukraine
\par}

\vspace{2mm}\par\noindent{\small\itshape
$^\ddag$~Department of Mathematics, Kyiv Academic University, Kyiv, Ukraine
\par}

\vspace{2mm}\par\noindent{\small\itshape
$^\S$~Borys Grinchenko Kyiv Metropolitan University, Kyiv, Ukraine
\par}

\vspace{5mm}\par\noindent{\small
E-mail:
\href{mailto:boyko@imath.kyiv.ua}{boyko@imath.kyiv.ua},
\href{mailto:sasha.lokazuik@gmail.com}{sasha.lokazuik@gmail.com}
\par}

\begin{abstract}
We revisit the group classification of the $(1{+}1)$-dimensional evolution
equations $u_t=F(u_n)$ and $u_t+uu_x=F(u_n)$ of arbitrary order $n\geq2$,
where $u_n=\partial_x^n u$, and describe the usual point equivalence groups
of the two classes.
We recover the known contact equivalences between the three
second-order equations with seven-dimensional maximal Lie invariance
algebras and extend the hodograph construction to points where $u_x=0$.
For $n>2$, the dimensions of the contact symmetry algebras
distinguish the two classes. We give reciprocal transformations
linking the two classes at order three and formulate a nonlocal
classification problem.
We also classify the extensions of the Lie invariance algebra of the
Burgers equation in a~wider class of Galilei-invariant equations.
\end{abstract}

\noindent\textbf{Keywords:} group classification, Lie symmetries,
contact transformations, equivalence transformations, evolution equations.

\noindent\textbf{2020 Mathematics Subject Classification:}
35A30, 35Q53, 58J70.

\bigskip

Consider the class of $(1{+}1)$-dimensional evolution equations
\begin{equation}\label{main-class}
u_t=\mathcal F(t,x,u,u_1,\dots,u_n),\qquad
n\geq2,\quad \mathcal F_{u_n}\ne0.
\end{equation}
Here $\mathcal F$ is smooth. We write $D_t$ and $D_x$ for the total
derivatives with respect to~$t$ and~$x$, respectively, and set $u_i=D_x^i u$
for $i\geq0$, so $u_0=u$ and $p=u_1=u_x$. On solutions of~\eqref{main-class},
we have $D_tu_i=D_x^i\mathcal F$.
We denote the maximal Lie algebra of point symmetries by $\mathfrak g$
and the maximal contact symmetry algebra by~$\mathfrak c$.
Unless otherwise specified, we work over~$\mathbb R$ on local domains
where the displayed powers are smooth.

The local contact transformations preserving the class~\eqref{main-class}
are of the form \cite[equations~(0.2) and~(0.3)]{maga1993a}
\[
(\widetilde t,\widetilde x,\widetilde u)
 =(T(t),\phi(t,x,u,p),\psi(t,x,u,p)),\qquad
 \phi_p\,\mathcal D\psi=\psi_p\,\mathcal D\phi.
\]
Here $\mathcal D=\partial_x+p\partial_u$, so $D_x=\mathcal D+u_2\partial_p$
on functions of $(t,x,u,p)$. The transformed first derivative is
$\widetilde u_{\widetilde x}=D_x\psi/D_x\phi$.
We require $T_tD_x\phi\ne0$ and local invertibility of the mapping
$(x,u,p)\mapsto(\widetilde x,\widetilde u,\widetilde u_{\widetilde x})$
at fixed~$t$.

The characteristic $h=A(t)u_t+B(t,x,u,p)$ defines the contact
vector field \cite[equations~(0.8) and~(0.9)]{maga1993a}
\[
\begin{gathered}
Q=\tau\partial_t+\xi\partial_x+\eta\partial_u+
  \zeta\partial_{u_t}+\rho\partial_p,\\
(\tau,\xi,\eta,\zeta,\rho)
 =(-A,-B_p,B-pB_p,h_t+u_t h_u,\mathcal D h).
\end{gathered}
\]
Here $h_t$ and $h_u$ are partial derivatives on the first jet space,
where $u_t$ and $p$ are independent coordinates.
If $B$ is affine in $p$, then $Q$ is the first prolongation of a point
vector field.

In the local analytic setting over $\mathbb C$, Magadeev's
Theorems~0.1 and~3.5 \cite{maga1993a} imply that~$\mathfrak c$ is
infinite-dimensional precisely for equations that are contact-equivalent
to linear ones. For finite-dimensional $\mathfrak c$, we have
$\dim\mathfrak c\leq n+5$, with equality precisely for equations that are
contact-equivalent to $u_t=u_n^{k_n}$, where $k_n=(1-n)/(1+n)$
and $u_n\ne0$ locally. This representative also appears in Sokolov's
list \cite[\S~1, Example~4]{Soko1988}. Its maximal contact symmetry algebra,
in the characteristic representation, is
\cite[proof of Theorem~3.5 and Appendix]{maga1993a}
\begin{equation}\label{cont-algebra-Mag-Class}
\mathfrak c=\big\langle x^j,\ p,\ xp-\tfrac{n-1}{2}u,\
 x^2p-(n-1)xu,\ u_t,\ tu_t-\tfrac{n+1}{2n}u\big\rangle,\quad j=0,\ldots,n-1.
\end{equation}
Each of these characteristics is affine in $u_t$ and $p$, so
$\mathfrak c=\operatorname{pr}^{(1)}\mathfrak g$.

We first consider the point symmetries of the subclass of \eqref{main-class}
\begin{equation}\label{eq:h-class}
u_t=F(u_n),\qquad n\geq2,\quad F'(u_n)\ne0.
\end{equation}

For $n=2$, Akhatov, Gazizov and Ibragimov carried out the group
classification of the nonlinear equations in~\eqref{eq:h-class} and
described the point equivalence transformations
\cite[equation~(3.14) and \S~4]{akha1991a}. Huraka and Lokaziuk gave an
algebraic proof for $n=2$ \cite{HuraLoka2024}.
Tsyfra studied nonlocal reductions of $u_2=F^{-1}(u_t)$
\cite[equations~(1)--(3)]{Tsyfra1995}.
For $n\geq3$, we use Ichanska's results
\cite[Theorems~7--9]{Icha2006} for the implicit form $u_n=F^{-1}(u_t)$.
Sepanski~\cite{Sepa2009} studied global actions of the symmetry groups
of nonlinear second-order equations from this class.

The usual point equivalence group $G^\sim_{\mathrm I}$ of
the class~\eqref{eq:h-class}
consists of the transformations
\[
\widetilde t=at+t_0,\quad \widetilde x=bx+x_0,\quad
\widetilde u=cu+dt+P_n(x),\quad abc\ne0,
\]
where $P_n$ is an arbitrary polynomial of degree at most $n$.
The induced action on the arbitrary element $F$ is
\begin{equation}\label{eq:affine-action}
\widetilde F(\widetilde u_n)
 =\delta_1F(\delta_2\widetilde u_n+\delta_3)+\delta_0,\qquad
 \delta_1\delta_2\ne0.
\end{equation}
Here $(\delta_1,\delta_2,\delta_3,\delta_0)
=(c/a,b^n/c,-P_n^{(n)}/c,d/a)$.
The real parameters $\delta_i$ are independent subject to $\delta_1\delta_2\ne0$,
and $\widetilde u_n=\partial_{\widetilde x}^{\,n}\widetilde u$.
The logarithmic and nonlinear power representatives may be taken on $u_n>0$.

For $n>2$, the contact equivalence group of the class~\eqref{eq:h-class}
consists of the first prolongations of the above point transformations.
For $n=2$, it consists of the transformations
\[
\begin{gathered}
\widetilde t=at+t_0,\qquad
\widetilde x=\alpha x+\beta p+x_0,\qquad
\widetilde p=\gamma x+ \sigma p+p_0,\\
\widetilde u=\Delta u+\tfrac12\alpha\gamma x^2+\beta\gamma xp+
 \tfrac12\beta \sigma p^2+p_0(\alpha x+\beta p)+dt+u_\ast,\\
\widetilde u_2=\frac{\gamma+\sigma u_2}{\alpha+\beta u_2},\qquad
\widetilde F(\widetilde u_2)=\frac{\Delta}{a}F(u_2)+\frac da,
\qquad \alpha+\beta u_2\ne0.
\end{gathered}
\]
Here $\Delta=\alpha\sigma-\beta\gamma$, $a\Delta\ne0$ and
$\widetilde p=\widetilde u_{\widetilde x}$. All transformation parameters
are real. The point equivalence subgroup is specified by $\beta=0$.

To see necessity, put $(X,U,P)=(\widetilde x,\widetilde u,\widetilde p)$,
$M=U_u-PX_u$ and $N=U_t-PX_t$. Then
$\widetilde u_{\widetilde t}=(N+Mu_t)/T_t$. Splitting with respect
to $F$ and $F'$ on the level sets of $\widetilde u_n$ gives
$\widetilde u_n=K(u_n)$ and constant ratios $M/T_t\ne0$ and $N/T_t$.
For $n>2$, $\partial_{u_n}\widetilde u_n=M/(D_xX)^n$ is constant,
so $X_p=U_p=0$. For a point transformation, $M/T_t$ being independent
of $p$ forces $X=X(t,x)$ and $U=M(t)u+U_0(t,x)$.
The condition on $N/T_t$ gives $X_t=M_t=0$ and $(U_0)_t=\mathrm{const}$.
The slope and intercept of $K$ then give $X_x=\mathrm{const}$ and
$\partial_x^n U_0=\mathrm{const}$.

For $n=2$, the coefficient matrix of
$K(u_2)=(\mathcal DP+u_2P_p)/(\mathcal DX+u_2X_p)$ is proportional
to a constant matrix and has determinant $M$. Since $M/T_t$ is constant,
the proportionality factor depends only on $t$. The spatial contact
identity $\mathrm dU-P\,\mathrm dX=M(\mathrm du-p\,\mathrm dx)$ then
gives $M_t=X_t=P_t=0$ upon differentiation in $t$.
Integrating the identity yields the stated transformations.
In particular, they include the lifts of the potential transformations
in \cite[equations~(6) and~(7)]{PopovychIvanova2005}, namely
$(x+\varepsilon p,u+\varepsilon p^2/2,p)$ and $(p,xp-u,x)$ in $(x,u,p)$.

\begin{theorem}\label{thm-h-classification}
The kernel Lie invariance algebra of the class~\eqref{eq:h-class}
has dimension $n+3$ and is given by
\[
\mathfrak g^\cap_{\mathrm I}=\langle \partial_t,\partial_x,S_n,\,
                  x^j\partial_u\ (j=0,\dots,n-1)\rangle,\qquad
S_n=nt\partial_t+x\partial_x+nu\partial_u.
\]
The table lists the kernel and all proper Lie symmetry extensions
up to $G^\sim_{\mathrm I}$.
In the row $F(u_n)=u_n^k$, we take
$k\in\mathbb R\setminus\{0,1,k_n\}$ and exclude $k=1/3$ for $n=2$.
\[\def\arraystretch{1.5}
\begin{array}{@{}l@{\qquad}l@{\qquad}c@{}}
 \text{\normalfont $F(u_n)$}
   & \mathfrak g(F) & \dim\mathfrak g(F) \\ \hline
 \text{\normalfont arbitrary $F$ (kernel)} & \mathfrak g^\cap_{\mathrm I} & n+3\\
 \mathrm{e}^{u_n} & \mathfrak g^\cap_{\mathrm I}+\langle E_n\rangle & n+4\\
 \ln u_n & \mathfrak g^\cap_{\mathrm I}+\langle L_n\rangle & n+4\\
 u_n^k & \mathfrak g^\cap_{\mathrm I}+\langle M^{(k)}\rangle & n+4\\
 u_n^{k_n},\quad k_n=\tfrac{1-n}{1+n}
   & \mathfrak g^\cap_{\mathrm I}+\langle M^{(k_n)},K_n\rangle & n+5\\
 u_2^{1/3},\quad n=2 & \mathfrak g^\cap_{\mathrm I}+\langle M^{(1/3)},R\rangle & 7\\
 u_n & \mathfrak g_{\rm lin} & \infty
\end{array}
\]
The additional generators for nonlinear $F$ are
\begin{gather*}
E_n=t\partial_t-\tfrac{x^n}{n!}\partial_u,\qquad
L_n=x\partial_x-nt\partial_u,\qquad
M^{(k)}=(1-k)t\partial_t+u\partial_u,\\
K_n=x^2\partial_x+(n-1)xu\partial_u,\qquad R=u\partial_x.
\end{gather*}
For affine $F$, \eqref{eq:affine-action} gives the representative $F(u_n)=u_n$.
Let $\mathcal I_n=\{\chi(t,x)\partial_u\mid\chi_t=\partial_x^n\chi\}$ be
the ideal of linear superposition symmetries.
For $n\geq3$, the classification in
\textnormal{\cite[Theorem~15, Table~1, Case~5]{BihloPopovych2017}} gives
\[
\mathfrak g_{\rm lin}
 =\mathcal I_n+\langle\partial_t,\partial_x,S_n,u\partial_u\rangle.
\]
For $n=2$, the displayed span is augmented by
$2t\partial_x-xu\partial_u$ and
$4t^2\partial_t+4tx\partial_x-(x^2+2t)u\partial_u$.
\end{theorem}

\begin{remark}
The equation with exponent $k_n$ is Magadeev's representative
attaining the bound $n+5$. For $n=2$, we have $k_2=-1/3$. The exponent
$1/3$ follows from the invariance condition $3k-1=0$ for $R$.
Both equations have maximal Lie invariance algebras of dimension seven
\cite[\S~4.3]{akha1991a} and occur up to point equivalence in the
classification by Zhdanov and Lahno \cite[equations~(4.9) and~(4.10),
with their arbitrary function set equal to~$1$]{ZhdanovLahno2007}.
In their variables, the maps to $u_t=u_2^{1/3}$ and
$u_t=u_2^{-1/3}$ are, respectively, $(t,x,xu)$ and $(t,u,-1/x)$
in $(\widetilde t,\widetilde x,\widetilde u)$, with $xu_x\ne0$ for the latter.
\end{remark}

We next consider the subclass of~\eqref{main-class} given by
\begin{equation}\label{eq:galilei-class}
u_t+uu_x=F(u_n),\qquad n\geq2,\quad F'(u_n)\ne0.
\end{equation}
Every equation in this class is invariant under the Galilei group $G(1,1)$,
generated by translations in $t$ and $x$ and the Galilei boosts
$(t,x,u)\mapsto(t,x+vt,u+v)$, $v\in\mathbb R$. Each boost gives
$\widetilde u_{\widetilde t}=u_t-vu_x$, $\widetilde u_{\widetilde x}=u_x$
and $\widetilde u_n=u_n$. Hence
\[
\widetilde u_{\widetilde t}+\widetilde u\,\widetilde u_{\widetilde x}
=u_t+uu_x.
\]

The group classification was carried out for $n=2,3,4$ in
\cite[Theorems~1--3]{Fush1996} and
\cite[Theorems~1--3]{Boyk1997} and for arbitrary $n\geq2$ in
\cite[Theorems~1--3]{Boyk2001}. An algebraic proof for $n=2$ is given in
\cite[Theorems~3.1--3.3 and Table~3.1]{HuraLoka2024}.
The usual point equivalence group $G^\sim_{\mathrm{II}}$ consists of the transformations
\[
\widetilde t=at+t_0,\quad
\widetilde x=bx+\tfrac12 ac t^2+avt+x_0,\quad
\widetilde u=\tfrac ba u+ct+v,
\]
where $ab\ne0$. Their action on $F$ is given by~\eqref{eq:affine-action}
with $(\delta_1,\delta_2,\delta_3,\delta_0)=(b/a^2,ab^{n-1},0,c/a)$.

The same splitting argument for point transformations gives
constant $M/T_t\ne0$ and $(N+T_tUP-Mup)/T_t$.
With $U=M(t)u+U_0(t,x)$, the coefficients of $up$ and $p$ give
$X_x=T_tM$ and $X_t=T_tU_0$. Constancy of $M/X_x^n$ then forces
$T_t$ and $M$ to be constant. The remaining condition gives
$U_0=ct+v$, yielding the transformation above.

\begin{theorem}\label{thm-galilei-classification}
The kernel Lie invariance algebra of the class~\eqref{eq:galilei-class}
is $\mathfrak g^\cap_{\mathrm{II}}=\langle\partial_t,\partial_x,G\rangle$,
where $G=t\partial_x+\partial_u$. The table lists the kernel
and all proper Lie symmetry extensions up to $G^\sim_{\mathrm{II}}$.
In the row $F(u_n)=u_n^k$, we take
$k\in\mathbb R\setminus\{0,3/(n+1)\}$ and exclude $k=1/3$ for $n=2$.
\[\def\arraystretch{1.5}
\begin{array}{@{}l@{\qquad}l@{\qquad}c@{}}
 \text{\normalfont $F(u_n)$}
   & \mathfrak g(F) & \dim\mathfrak g(F) \\ \hline
 \text{\normalfont arbitrary $F$ (kernel)} & \mathfrak g^\cap_{\mathrm{II}} & 3\\
 \ln u_n & \mathfrak g^\cap_{\mathrm{II}}+\langle D_n^{(\ln)}\rangle & 4\\
 u_n^k & \mathfrak g^\cap_{\mathrm{II}}+\langle D_n^{(k)}\rangle & 4\\
 u_n^{3/(n+1)} & \mathfrak g^\cap_{\mathrm{II}}+\langle H,\Pi\rangle & 5\\
 u_2^{1/3},\quad n=2 & \mathfrak g_{*} & 7
\end{array}
\]
The additional generators are
\begin{align*}
D_n^{(k)}&=\big((n-1)k+1\big)t\partial_t+(2-k)x\partial_x+
                  (1-nk)u\partial_u,\\
D_n^{(\ln)}&=t\partial_t+\big(2x-\tfrac{2n-1}{2}t^2\big)\partial_x
                         +\big(u-(2n-1)t\big)\partial_u,\\
H&=2t\partial_t+x\partial_x-u\partial_u,\qquad
\Pi=t^2\partial_t+tx\partial_x+(x-tu)\partial_u.
\end{align*}
For $n=2$ and $F(u_2)=u_2^{1/3}$, the maximal Lie invariance algebra is
\begin{equation}\label{LieSym-BF}
\mathfrak g_{*}=\mathfrak g^\cap_{\mathrm{II}}
 +\langle 3D_2^{(1/3)},\,R,\,(2tu-x)\partial_x+u\partial_u,\,(tu-x)G\rangle,
\end{equation}
where $R$ is defined in Theorem~\textup{\ref{thm-h-classification}}.
\end{theorem}

For both classes, write $u_t+\mu up=F(u_n)$, $\mu\in\{0,1\}$.
The contact determining equation for $h=A(t)u_t+B(t,x,u,p)$ is
\begin{equation} \label{eq:common-determining}
(A_t+B_u)F+B_t+\mu(uB_x+pB-p^2B_p-A_tup)
=F'\bigl(D_x^nB-B_pu_{n+1}\bigr).
\end{equation}
For point symmetries take $A=-\tau$ and $B=\eta-\xi p$, so the
expression in parentheses on the right is the spatial prolongation
coefficient $\eta^{(n)}$. For $n\geq3$, the coefficient of
$u_2u_{n-1}$ gives $\xi_u=0$, using $u_2^2$ when $n=3$.
The coefficient of $u_{n-1}$ then gives $\eta=au+b$ and
$2a_x=(n-1)\xi_{xx}$, where $a,b$ depend on $(t,x)$.

The tables in Theorems~\ref{thm-h-classification}
and~\ref{thm-galilei-classification} follow by furcate splitting
\cite[arXiv version, \S~5, after equation~(8)]{NikitinPopovych2001}.
For non-affine $F$, the space of constant-coefficient relations
$(a_1u_n+a_0)F'+b_1F=b_0$ has dimension at most one. Indeed,
two independent relations would yield a nonzero one with $a_1+b_1=0$.
Differentiating it gives $(a_1u_n+a_0)F''=0$, which forces
all four coefficients to vanish.

For class~\eqref{eq:h-class}, splitting \eqref{eq:common-determining}
for non-affine $F$ gives $\xi_t=a_t=0$, $\xi_{xxx}=0$ and
\[
\bigl((a-n\xi_x)u_n+\partial_x^n b\bigr)F'+(\tau_t-a)F=b_t.
\]
Its nontrivial cases reduce by~\eqref{eq:affine-action} to the exponential,
logarithmic and power representatives. If $\xi_{xx}\ne0$, differentiation
in $x$ fixes the power at $k_n=(1-n)/(1+n)$.

For class~\eqref{eq:galilei-class}, splitting gives
$\xi=\lambda(t)x+\sigma(t)$, $a=\lambda-\tau_t$, $b=\xi_t$,
$2\lambda_t=\tau_{tt}$ and $\lambda_{tt}=0$. The classifying equation is
\[
 \bigl((n-1)\lambda+\tau_t\bigr)u_n F'
   +(\lambda-2\tau_t)F=-\sigma_{tt}.
\]
The coefficient multiplying $F'$ is proportional to $u_n$, giving
logarithmic and unshifted power cases for non-affine $F$.
If $\lambda_t\ne0$, differentiation in $t$ fixes the power at $3/(n+1)$
after an additive normalization of $F$. The preceding regular splittings
also hold at $n=2$ when $\xi_u=0$.

For $n=2$ with $\xi_u\ne0$, the coefficient of $p$ in
\eqref{eq:common-determining} gives $F=c_1(u_2+r_0)^{1/3}+c_0$,
$c_1\ne0$. The shift $r_0$ can be removed in~\eqref{eq:h-class}.
In~\eqref{eq:galilei-class}, first remove $c_0$ and write
$\xi=\alpha u+\beta$ and $\eta=\alpha_xu^2+au+b$, as given by the
$p^3$ and $p^2$ coefficients. The equation independent of $p$ gives
$\eta_t+u\eta_x=0$ and $2\eta_u+2\xi_x=3\tau_t$, hence
$\alpha_x=a_x=\beta_{xx}=0$. The $F'$ term in the coefficient of $p$
now gives $r_0=0$. The remaining equations give the basis in~\eqref{LieSym-BF}.

For affine $F$, the normal form $u_t+uu_x=u_n$ is in the row
with $k=1$ for $n>2$ and in the row with exponent $3/(n+1)$ for $n=2$.
These cases also occur in \cite[Table~1, Case~11]{OpanBihloPopovych2017}.

Theorems~\ref{thm-h-classification} and~\ref{thm-galilei-classification}
give the following nonlinear second-order equations with maximal Lie
invariance algebras of dimension seven:
\begin{align}
u_t&=u_2^{-1/3},\label{-1/3}\\
u_t&=u_2^{1/3},\label{1/3}\\
u_t+uu_x&=u_2^{1/3}.\label{eq_Boyko}
\end{align}
We denote their maximal Lie invariance algebras by $\mathfrak g_-$,
$\mathfrak g_+$ and $\mathfrak g_{*}$, respectively. We work locally
with $u_2\ne0$ and use the real cube root. Equation~\eqref{1/3} also
describes the affine heat flow of plane curves in graph coordinates
\cite[equation~(33)]{AngenentSapiroTannenbaum1998}.

The Legendre transformation \cite{Gazi1994,Pukh1996} combined with
time reversal maps~\eqref{1/3} to~\eqref{-1/3}
\cite[equation~(9)]{boyk2019a},
\begin{equation}\label{Legendre}
(\widetilde t,\widetilde x,\widetilde u)=(-t,u_x,xu_x-u),\qquad
  (\widetilde u_{\widetilde t},\widetilde u_{\widetilde x},
   \widetilde u_2)=(u_t,x,1/u_2).
\end{equation}
With $\widetilde t=t$, the Legendre transformation gives
$\widetilde F(\widetilde u_2)=-F(1/\widetilde u_2)$
\cite[p.~345]{King1992}. It is also the lift to $u$ of the potential hodograph
transformation \cite[equation~(7)]{PopovychIvanova2005} for
$r_t=D_x(F'(r)r_x)$ with $r=u_2$ and potential $p=u_x$.

For $u_x\ne0$, the usual and modified hodograph transformations
\cite[equation~(1)]{FushTych1993}, \cite[p.~20, equation~(10)]{boyk2019a}
are, respectively,
\[
(t,x,u)\mapsto(t,u,x),\qquad (t,x,u)\mapsto(t,u,x-tu).
\]
Both transformations belong to the family of point transformations
\begin{equation}\label{local1}
(\widetilde t,\widetilde x,\widetilde u)
  =\bigl(t,u+\varepsilon(x-\mu tu),x-\mu tu\bigr),
  \qquad \mu\in\{0,1\},\quad \varepsilon\in\mathbb R.
\end{equation}
For $\Lambda=D_x\widetilde x=\varepsilon+(1-\varepsilon\mu t)u_x\ne0$,
we obtain
\[
\widetilde u_{\widetilde t}=-\frac{u_t+\mu uu_x}{\Lambda},\qquad
\widetilde u_{\widetilde x}=\frac{1-\mu t u_x}{\Lambda},\qquad
\widetilde u_2=-\frac{u_2}{\Lambda^3}.
\]
For $\mu=0$, transformation~\eqref{local1} preserves~\eqref{1/3}.
For $\mu=1$, it maps~\eqref{eq_Boyko} to~\eqref{1/3}.
Setting $\varepsilon=0$ gives the two hodograph transformations above.
The Jacobian with respect to $(x,u)$ is $-1$. At a point where
$u_x=0$, choosing $\varepsilon=1$ gives $\Lambda=1$, so the transformed
graph is locally regular.

\begin{remark}
For $n=2$ and $u_x\ne0$, the modified hodograph transformation maps~\eqref{eq:galilei-class} to
\[
\widetilde u_{\widetilde t}
=-(\widetilde t+\widetilde p)\,
 F\!\left(-\frac{\widetilde u_2}{(\widetilde t+\widetilde p)^3}\right),
\qquad \widetilde p=\widetilde u_{\widetilde x}.
\]
Independence of the right-hand side from $\widetilde p$ is equivalent to
$3u_2F'(u_2)=F(u_2)$. Thus the image belongs to~\eqref{eq:h-class} precisely
for $F(u_2)=\kappa u_2^{1/3}$, $\kappa\ne0$, in which case
$\widetilde F(\widetilde u_2)=\kappa\widetilde u_2^{1/3}$.
For $u_t+c(u)u_x=0$, the characteristic formula $x-tc(u)=f(u)$
\cite[equations~(2.5) and~(2.6)]{Whitham1974} gives the reduction to
$\widetilde u_{\widetilde t}=0$ on $u_x\ne0$ via
$(t,x,u)\mapsto(t,u,x-tc(u))$. The choice $c(u)=u$ gives the modified
hodograph transformation. For $u_t+c(t,u)u_x=0$, the usual hodograph
transformation gives $\widetilde u_{\widetilde t}=c(\widetilde t,\widetilde x)$
\cite[following equation~(13)]{FushTych1993}.
\end{remark}

\begin{remark}
The determining equation for a contact symmetry of~\eqref{-1/3} is
\[
\mathcal D^2 B+(3A_t+4B_u+2\mathcal D B_p)u_2
+B_{pp}u_2^2+3B_tu_2^{4/3}=0.
\]
On either half-axis $u_2\ne0$, the functions $1,u_2,u_2^2,u_2^{4/3}$
are linearly independent. Splitting gives
$B=\beta(x)p-(\tfrac12\beta_x+\tfrac34 A_t)u+\gamma(x)$ with
$A_{tt}=\beta_{xxx}=\gamma_{xx}=0$. The seven integration constants give
the characteristics in~\eqref{cont-algebra-Mag-Class} for $n=2$, so the
maximal contact symmetry algebra of~\eqref{-1/3} is
$\operatorname{pr}^{(1)}\mathfrak g_-$.
Composing~\eqref{local1} with $\mu=1$ and~\eqref{Legendre} maps
\eqref{eq_Boyko} to Magadeev's representative~\eqref{-1/3}. Thus the three
maximal contact symmetry algebras are isomorphic and attain Magadeev's
bound $2+5=7$. The maximal contact symmetry algebra of each
equation is the first prolongation of its maximal Lie invariance algebra.
\end{remark}

For $n=2$, the representative $F(u_2)=u_2$ in~\eqref{eq:galilei-class}
is the Burgers equation. Its maximal Lie invariance algebra is
$\mathfrak b=\mathfrak g^\cap_{\mathrm{II}}+\langle H,\Pi\rangle$.
The Cole--Hopf transformation \cite{Cole1951,Hopf1950} linearizes
$u_t+uu_x=\nu u_2$ for $\nu\ne0$,
\begin{equation}\label{eq:cole-hopf}
u=-2\nu(\ln|v|)_x,\qquad v_t=\nu v_{xx},\qquad v\ne0.
\end{equation}
To recover $v$ from $u$, integrate $(\ln|v|)_x=-u/(2\nu)$
and choose the time-dependent factor so that $v_t=\nu v_{xx}$.
The Legendre transformation with $\widetilde t=t$ linearizes
$u_t=-1/u_2$, so~\eqref{eq:cole-hopf} also links that equation to the Burgers
equation with $\nu=1$. Its differential consequence $p_t=p_x^{-2}p_{xx}$
is linearized by the usual hodograph transformation
\cite[\S~1, equations~(1)--(3) and Example~1]{FushTych1993}.

Fushchych and Boyko \cite[equations~(28) and~(31)]{Fush1996}
and Boyko \cite[equations~(20) and~(21)]{Boyk1997} considered
the following class and its finite-sum subclass.
\begin{align}
u_t+uu_x &= u_2\Phi(\omega_3,\ldots,\omega_n),\qquad
\omega_j=\frac{u_j^{3/(j+1)}}{u_2},\quad j=3,\ldots,n,\label{eq:boyko-generalizations1}\\
u_t+uu_x &= \sum_{j=2}^{n}\lambda_j u_j^{3/(j+1)}.\label{eq:boyko-generalizations2}
\end{align}
Here $\Phi$ is smooth and the $\lambda_j$ are constants.
On $u_2\ne0$, the choice
$\Phi=\lambda_2+\sum_{j=3}^n\lambda_j\omega_j$ yields
\eqref{eq:boyko-generalizations2} from~\eqref{eq:boyko-generalizations1}.
All these equations admit $\mathfrak b$ by Theorem~8 in both cited works.
Both classes contain the equation with right-hand side $u_n^{3/(n+1)}$,
whose maximal Lie invariance algebra is~$\mathfrak b$ by
Theorem~\ref{thm-galilei-classification}. Thus their kernel is
the five-dimensional Lie invariance algebra~$\mathfrak b$ of the Burgers equation.
For~\eqref{eq:boyko-generalizations2}, this algebra is maximal if some~$\lambda_j$ is nonzero \cite[p.~127]{Boyk1997}.

The choices $\Phi=\nu\ne0$ and $\lambda_2=\nu\ne0$,
$\lambda_3=\dots=\lambda_n=0$, respectively, give the Burgers equation.
If $\Phi\equiv0$ or all $\lambda_j$ vanish, the equation is
first-order. The usual hodograph transformation gives
$\widetilde u_{\widetilde t}=\widetilde x$ where $u_x\ne0$, so its
maximal Lie invariance algebra is infinite-dimensional.

\begin{theorem}\label{thm-burgers-family}
Consider an equation of order $n$ from~\eqref{eq:boyko-generalizations1}
on a local domain with $u_2\ne0$ where the powers are smooth.
For $n=2$, $\Phi$ is a nonzero constant and the maximal Lie invariance
algebra is $\mathfrak b$. For $n\geq3$, assume $\Phi_{\omega_n}\ne0$
on this domain.
The maximal Lie invariance algebra is
$\mathfrak b+\langle Z\rangle$, where $Z=x\partial_x+u\partial_u$, if and only if
\[
\sum_{j=3}^n\frac{j-2}{j+1}\,\omega_j\Phi_{\omega_j}=-\Phi.
\]
For all other $\Phi$, the maximal Lie invariance algebra is~$\mathfrak b$.
The homogeneity equation has the following local general solution
on $\omega_n\ne0$.
\[
\Phi=|\omega_n|^{-\frac{n+1}{n-2}}
\Psi\!\left(
 \left(\omega_j|\omega_n|^{-\frac{(j-2)(n+1)}{(j+1)(n-2)}}\right)_{j=3}^{n-1}
\right).
\]
Here $\Psi$ is an arbitrary smooth function subject to
$\Phi_{\omega_n}\ne0$. For $n=3$, it is a nonzero constant.
\end{theorem}

\begin{proof}
Theorem~\ref{thm-galilei-classification} gives the assertion for $n=2$.
For $n\geq3$, the time component of a point symmetry generator is
$\tau(t)$ by the general form of contact generators
\cite[equation~(0.8)]{maga1993a}. The infinitesimal invariance criterion for
$Q=\tau(t)\partial_t+\xi\partial_x+\eta\partial_u$ gives
\[
0=\eta_t+u\eta_x+\bigl(\eta+u(\tau_t-\xi_x)-\xi_t\bigr)p
 +(\eta_u-\tau_t-p\xi_u)u_2\Phi
 -\sum_{j=2}^n\eta^{(j)}\partial_{u_j}(u_2\Phi),
\]
where $\xi$ and $\eta$ depend on $(t,x,u)$ and
$\eta^{(j)}=D_x^j\eta-\sum_{i=1}^j\binom{j}{i}u_{j+1-i}D_x^i\xi$.
The coefficient of~$p^{n+1}$ gives $\partial_u^n\xi=0$ since
$\partial_{u_n}(u_2\Phi)\ne0$.
All terms involving $\xi$ in the coefficient of $p^n$ contain
at least $n-1$ derivatives with respect to $u$. Differentiating this
coefficient in $u$ therefore gives $\partial_u^{n+1}\eta=0$.
Let $m=\deg_u\xi$ if $\xi\ne0$ and $m=0$ otherwise.
The coefficient of~$p$ gives $\deg_u\eta\leq m+1$ since the other terms
in this coefficient either involve a derivative of~$\eta$ with respect
to $u$ or have degree at most $m+1$ in $u$.

Assign weight $j+1$ to $u_j$ for $j\geq2$ and weight zero to $t$, $x$, $u$, $p$.
Each $\omega_j$ has weight zero, so $u_2\Phi$ and
$\partial_{u_j}(u_2\Phi)$ have weights $3$ and $2-j$, respectively.
Suppose $m>0$. Since $\deg_u\eta\leq m+1$, only derivatives
$\partial_u^r\eta$ with $r\leq2$ and their $x$-derivatives can contribute
in $D_x^j\eta$ to the coefficient of $u^{m-1}p$. One factor is $p$, so
there can be at most one further jet factor $u_i$ with $2\leq i\leq j-1$.
Thus these terms have weight at most $j$. For $\xi$, only derivatives
with at most one $u$-differentiation can contribute at degree $m-1$.
The terms in $\eta^{(j)}$ of weight $j+1$ contributing to the coefficient
of $u^{m-1}p$ therefore come from $-(j+1)\xi_u p u_j$.
The terms of weight $3$ in the determining equation at $u^{m-1}p$
then come from
\[
\xi_u p\left(\sum_{j=2}^n(j+1)u_j\partial_{u_j}(u_2\Phi)-u_2\Phi\right)
 =2\xi_u p u_2\Phi.
\]
Their vanishing gives $m=0$. Thus $\xi=\xi(t,x)$ and
$\eta=a_1(t,x)u+a_0(t,x)$.

The coefficients of $up$ and $u^2$ give
$a_1=\xi_x-\tau_t$ and $(a_1)_x=0$,
so $\xi_{xx}=0$. The coefficient of $p$ then gives
$a_0=\xi_t$ and $\eta^{(j)}=(a_1-j\xi_x)u_j$ for $j\geq2$.
The terms of weight zero give $\xi_{tt}=0$ and $2\xi_{tx}=\tau_{tt}$.
Hence $Q\in\mathfrak b+\langle Z\rangle$.

The flow of $Z$ scales $(x,u)$ by a common factor $\lambda>0$.
It scales $u_t+uu_x$, $u_2$ and~$\omega_j$ by~$\lambda$,~$\lambda^{-1}$
and $\lambda^{-2(j-2)/(j+1)}$, respectively. Differentiating the resulting
invariance condition at $\lambda=1$ gives the stated equation for~$\Phi$.
Integrating along characteristics yields the formula on $\omega_n\ne0$.
\end{proof}

For $n=3$ and $u_3>0$, the equations in~\eqref{eq:boyko-generalizations1}
with maximal Lie invariance algebra $\mathfrak b+\langle Z\rangle$ are
\[
u_t+uu_x=C\frac{u_2^5}{u_3^3},\qquad C\ne0.
\]
For any $n\geq3$, $\Phi=|\omega_n|^{-(n+1)/(n-2)}$ gives
\[
u_t+uu_x=\frac{u_2|u_2|^{(n+1)/(n-2)}}{|u_n|^{3/(n-2)}},
\qquad u_2u_n\ne0.
\]

The transformations generated by $\mathfrak b$, including translations,
Galilei boosts and projective transformations, fix~$\Phi$.
Every usual point equivalence transformation normalizes~$\mathfrak b$,
whose radical is $\langle\partial_x,G\rangle$ and whose time projection
is the projective algebra. Its time component is therefore fractional
linear. Projective transformations generated by~$\mathfrak b$ and
a time scaling in the normalizer reduce it to $T=t$. With $T=t$,
the action on the radical commutes with the standard~$\mathfrak{sl}_2$
action, so it is scalar. Hence $X=bx+X_0(t)$ and $U=bu+U_0(t)$.
Normalizing~$\partial_t$ and $H$ gives $U_0=v$ and $X_0=vt+x_0$.
Consequently, the action on $\Phi$ is represented by
$\widetilde t=(b/c)t$, $\widetilde x=bx$, $\widetilde u=cu$, $bc\ne0$,
on domains where the powers are defined. The induced action is
\[
\widetilde\Phi(\widetilde\omega_3,\ldots,\widetilde\omega_n)
 =bc\,\Phi(\omega_3,\ldots,\omega_n),\qquad
\widetilde u_j=\frac{c}{b^j}u_j,\qquad
\widetilde\omega_j=\frac{\widetilde u_j^{3/(j+1)}}{\widetilde u_2}.
\]
The coefficient $C$ in the third-order equation above is invariant under
this action.

For $n>2$, the classes~\eqref{eq:h-class} and~\eqref{eq:galilei-class}
are disjoint up to contact equivalence.
For $n>2$,
$\partial_{u_n}(D_x^nB-B_pu_{n+1})=B_u+n\mathcal D B_p+nu_2B_{pp}$.
Taking $\partial_{u_2}\partial_{u_n}$ in
\eqref{eq:common-determining} after division by $F'$ gives $nB_{pp}=0$.
Hence all contact symmetries are first prolongations of point symmetries.
Theorems~\ref{thm-h-classification} and~\ref{thm-galilei-classification}
give maximal contact symmetry algebras of dimension at least $n+3$
for~\eqref{eq:h-class} and at most $5$ for~\eqref{eq:galilei-class}.

For $u_t+uu_x=u_{xxx}$, the change $y=x-tu$, $v(t,y)=u(t,x)$
gives, where $1-tu_x\ne0$,
\[
v_t=\partial_y\left(\frac{v_{yy}}{(1+tv_y)^3}\right).
\]
This is the divergence form of \cite[Example~3]{PopovychSergyeyev2010}
with the sign of the dependent variable reversed.
For $n=3$, Tychynin studied nonlocal linearizations of equations of the
form $z_t=F'(z_{xx})z_{xxx}$, obtained from~\eqref{eq:h-class}
by $z=u_x$ \cite{Tychynin1994}.

The potential Harry Dym equation $u_t=u_3^{-1/2}$, $u_3>0$, is
Magadeev's representative for $n=3$, with $k_3=-1/2$. For this equation,
the substitution $w=u_3^{-1/2}$ gives $w_t=-\tfrac12 w^3w_{xxx}$
\cite[Example~9.2, equations~(9.17) and~(9.18)]{FelsYasar2019}.
In this normalization, the reciprocal link to the Korteweg--de~Vries
equation is given by \cite{HeremanEtAl1989,KonopelchenkoOevel1993}
\begin{equation}\label{eq:dym-kdv}
\mathrm{d}y=\frac{\mathrm{d}x}{w}+
 \left(\frac{ww_{xx}}2-\frac{w_x^2}4\right)\mathrm{d}t,\qquad
 v(t,y)=\frac{3}{4}\bigl(2ww_{xx}-w_x^2\bigr).
\end{equation}
The one-form defining $y$ is closed on solutions. In the inverse coordinate
$x=X(t,y)$, we have $X_t=-\tfrac12\{X;y\}X_y$ and $v=\tfrac32\{X;y\}$,
where $\{X;y\}$ is the Schwarzian derivative. Thus $v$ satisfies
$v_t+vv_y=-\tfrac12v_{yyy}$ in~\eqref{eq:galilei-class}
\cite[\S~3, Example~5]{Soko1988}.

The classification of integrable third-order equations in
\cite[\S~3.3]{MikShabSoko1991}, \cite{HernandezEtAl1995},
\cite[equation~(2.5)]{Hernandez2005} and
\cite[equation~(4.1c)]{BilgeMizrahi2017} includes, after rescaling $t$,
\begin{equation}\label{eq:quadratic-separant}
u_t=F(u_3),\qquad F'(u_3)=\mathcal P(u_3)^{-3/2},\qquad \mathcal P(u_3)>0,
\end{equation}
where $\mathcal P(u_3)=c_2u_3^2+c_1u_3+c_0$ and
$d_{\mathcal P}=c_1^2-4c_2c_0$.
We take $d_{\mathcal P}>0$ for the real defocusing reduction below.
The canonical density $\varrho=(F')^{-1/3}
=\sqrt{\mathcal P(u_3)}$ \cite[equation~(2.2)]{Hernandez2005}
gives the reciprocal transformation
\[
\mathrm{d}y=\varrho\,\mathrm{d}x+\mathcal J\,\mathrm{d}t,\qquad
\mathcal J=\varrho_{u_3}D_x^2F(u_3)+\tfrac18 d_{\mathcal P}\bigl(D_xF(u_3)\bigr)^2.
\]
The identity $\varrho_{u_3u_3}=-\tfrac14 d_{\mathcal P} F'(u_3)$ gives
$D_t\varrho=D_x\mathcal J$ along solutions in $(t,x)$.

In the reciprocal coordinates $(t,y)$, the function $\theta(u_3)$ with
$\theta'(u_3)=1/\mathcal P(u_3)$ satisfies $\theta_y=D_xF(u_3)$ and
$\theta_t=\theta_{yyy}-\tfrac18 d_{\mathcal P}\theta_y^3$.
Thus $q=\tfrac14\sqrt{d_{\mathcal P}}\,\theta_y$ satisfies the defocusing modified
Korteweg--de~Vries equation $q_t=q_{yyy}-6q^2q_y$, cf.~\cite{Kawamoto1985}
for the Harry Dym case. The Miura substitution $v=6(q_y+q^2)$ yields
$v_t+vv_y=v_{yyy}$ \cite{Miura1968}.
For $c_2<0$ and $c_2>0$, respectively, the normal forms
of~\eqref{eq:quadratic-separant} with $d_{\mathcal P}>0$ are
\[
u_t=\frac{u_3}{\sqrt{1-u_3^2}},\quad |u_3|<1,\qquad
u_t=-\frac{u_3}{\sqrt{u_3^2-1}},\quad |u_3|>1.
\]
The usual point equivalence group $G^\sim_{\mathrm I}$
preserves both the degree and the sign of the leading coefficient of
$\mathcal P$. The potential Harry Dym equation corresponds to linear
$\mathcal P$.

Scaling the preceding Korteweg--de~Vries equations gives
$v_t+vv_y+v_{yyy}=0$. For rapidly decaying solutions, inverse scattering
gives an integral link with the Airy equation \cite{GardnerEtAl1967}.
The input kernel $\Omega(t,z)$ in the right Marchenko equation
for the potential $-v/6$ satisfies the Airy equation
$\Omega_t+\Omega_{zzz}=0$
\cite[equations~(17)--(19), (23) and~(24)]{Tanaka1974}.

The Harry Dym and Korteweg--de~Vries hierarchies are reciprocally related
\cite{KonopelchenkoOevel1993,RogersNucci1986}.
The terms $vv_{yyy}$ and $v_yv_{yy}$ in the fifth-order
Korteweg--de~Vries flow place it outside~\eqref{eq:galilei-class}.

\begin{problem}%\label{prob:reciprocal-links}
For each fixed $n>2$, classify pairs of equations
from~\eqref{eq:h-class} and~\eqref{eq:galilei-class} such that a reciprocal
transformation followed by a finite-order differential substitution maps
the first equation to the second. Take the pairs up to
$G^\sim_{\mathrm I}$ and $G^\sim_{\mathrm{II}}$ in the respective classes.

Allow $\widetilde t=at+t_0$, $a\ne0$, and
$\mathrm{d}y=\varrho\,\mathrm{d}x+\mathcal J\,\mathrm{d}t$,
where $\varrho$ and $\mathcal J$ are local differential functions of
finite order on the source equation, $\varrho\ne0$ and
$D_t\varrho=D_x\mathcal J$. The substitution for $v(\widetilde t,y)$
may depend on $\widetilde t$, $y$, $X(\widetilde t,y)=x$,
$U(\widetilde t,y)=u$ and their $y$-derivatives. Its differential with
respect to the free dependent-variable jet coordinates must be nonzero
on the reciprocal system. Allow $\mathrm{d}y=\mathrm{d}x$ and use the
actual pair $(\varrho,\mathcal J)$ in testing the link.
Do such pairs occur at any order $n\geq4$?
\end{problem}

\textbf{Acknowledgments.}
V.M.B. gratefully acknowledges support from the National Research Foundation
of Ukraine through project 2025.07/0405, \emph{Algebraic methods for studying
equations of mathematical physics}. The contents of this work are the
sole responsibility of the Institute of Mathematics of the National
Academy of Sciences of Ukraine and do not necessarily reflect the views
of the National Research Foundation of Ukraine.
The authors are grateful to Roman Popovych and Oleksandra Vinnichenko
for helpful discussions and suggestions.

\end{document}